\documentclass[twocolumn]{emulateapj}
\usepackage{epsfig}

\newcommand\cmc{\mbox{cm$^{-2}$}}

\def\lsim {$\rlap{\raise.4ex\hbox{$<$}}\lower.55ex\hbox{$\sim$}\,$}

\newcommand{\iso}{\mbox{\it ISO}}
\newcommand{\spitzer}{{\it Spitzer}}

\newcommand{\water}{H$_2$O}
\newcommand{\ammonia}{\mbox{{\rm NH}$_3$}}
\newcommand{\coo}{CO$_2$}

\newcommand{\methane}{CH$_4$}
\newcommand{\ammonium}{\mbox{{\rm NH}$_4^+$}}
\newcommand{\methanol}{CH$_3$OH}
\newcommand{\nn}{N$_2$}
\newcommand{\ocn}{OCN$^{-}$}

\newcommand{\av}{$A_{\rm V}$}
\newcommand{\tsil}{$\tau_{9.7}$}

\slugcomment{To appear in Astrophysical Journal Letters}

\shorttitle{Ices Toward Background Stars}
\shortauthors{Knez et al.}

\begin{document}

\title{{\it Spitzer} Mid-infrared Spectroscopy of Ices toward Extincted
Background Stars}

\author{Claudia Knez\altaffilmark{1}}
\altaffiltext{1}{Department of Astronomy, University of Texas at
Austin, 1 University Station C1400, Austin, TX 78712-0259, USA;
claudia@astro.as.utexas.edu}

\author{A. C. Adwin Boogert\altaffilmark{2}}
\altaffiltext{2}{Division of PMA, Mail Code 105-24, California
Institute of Technology, Pasadena, CA 91125, USA}

\author{Klaus M. Pontoppidan\altaffilmark{3}}
\altaffiltext{3}{Division of GPS, Mail Code 150-21, California
Institute of Technology, Pasadena, CA 91125, USA}

\author{Jacqueline Kessler-Silacci\altaffilmark{1}}

\author{Ewine F. van Dishoeck\altaffilmark{4}}
\altaffiltext{4}{Leiden Observatory, PO Box 9513, 2300 RA Leiden,
The Netherlands}

\author{Neal J. Evans, II\altaffilmark{1}}

\author{Jean-Charles Augereau\altaffilmark{4,5}}
\altaffiltext{5}{Laboratoire d'Astrophysique de l'Observatoire de
Grenoble, B.P. 53, 38041 Grenoble Cedex 9, France}

\author{Geoffrey A. Blake\altaffilmark{3}}

\author{Fred Lahuis\altaffilmark{4,6}}
\altaffiltext{6}{SRON, PO Box 800, 9700 AV Groningen, The
Netherlands}

\begin{abstract}

A powerful way to observe directly the solid state inventory of
dense molecular clouds is by infrared spectroscopy of background
stars. We present \spitzer/IRS 5-20 \micron\ spectra of ices toward
stars behind the Serpens and Taurus molecular clouds, probing visual
extinctions of 10-34 mag. These data provide the first complete
inventory of solid-state material in dense clouds before star
formation begins. The spectra show prominent 6.0 and 6.85 \micron\
bands.  In contrast to some young stellar objects (YSOs), most
($\sim$75\%) of the 6.0 \micron\ band is explained by the bending
mode of pure \water\ ice. In realistic mixtures this number
increases to 85\%, because the peak strength of the \water\ bending
mode is very sensitive to the molecular environment.  The strength
of the 6.85 \micron\ band is comparable to what is observed toward
YSOs. Thus, the production of the carrier of this band does not
depend on the energetic input of a nearby source. The spectra show
large abundances of CO and CO$_2$ (20-40\% with respect to H$_2$O
ice). Compared to YSOs, the band profile of the 15 \micron\ CO$_2$
bending mode lacks the signatures of crystallization, confirming the
cold, pristine nature of these lines of sight. After the dominant
species are removed, there are residuals that suggest the presence
of minor species such as HCOOH and possibly \ammonia. Clearly,
models of star formation should begin with dust models already
coated with a fairly complex mixture of ices.

\end{abstract}

\section{Introduction}

Infrared absorption studies of protostars embedded in dense clouds
have shown that dust grains along these lines of sight have icy
mantles (e.g., Willner et al.\ 1982, Tielens et al.\ 1984,
Allamandola et al. 1992, Whittet et al.\ 1996, Boogert et al.\
2004). Heating by the protostar and energetic photons can affect the
ice composition by sublimation and by triggering chemical reactions
(e.g., Gerakines et al. 1996, Ehrenfreund \&\ Charnley 2000, Schutte
\& Khanna 2003). Knowledge of the ice composition in quiescent dense
clouds is required in determining the amount of processing ices
undergo during star formation.  This `baseline' can be obtained by
observations of field stars lying behind molecular clouds. These
observations also help constrain models of chemical evolution during
the star forming process (e.g., Lee et al.\ 2004).

The 3 \micron\ absorption band of \water\ ice was observed toward
stars behind the Serpens (Eiroa \&\ Hodapp 1989) and Taurus dark
clouds (Whittet et al.\ 1988, Smith et al.\ 1993, Murakawa et al.\
2000).  These studies indicate that \water\ ice is formed deep in
the clouds, at visual extinctions $A_{\rm V}>3$ magnitudes (the ice
formation threshold). Solid CO was observed toward background stars
as well (e.g., Whittet et al.\ 1985, Chiar et al.\ 1994, 1995). Its
formation threshold is significantly larger (\av=6-15 mag)  due to
the lower sublimation temperature of solid CO.

Studies of ices toward background stars have been limited to bands
below 5 \micron\ because of telluric absorption and the fact that
stellar fluxes drop rapidly with increasing wavelength. Observations
of background stars have become increasingly feasible
with the {\it Infrared Space Observatory (ISO)} mission 
and, in particular, with the launch of the {\it Spitzer Space
Telescope} (Werner et al.\ 2004).  With \iso, solid \coo\ at 4.25
\micron\ was observed toward two Taurus background stars (Whittet et
al.\ 1998; Nummelin et al.\  2001), indicating that radiation from
nearby protostars is not required to form this species.  Recent
observations with {\it Spitzer} detected the \coo\ bending mode at
15 \micron\ toward background stars (Bergin et al.\ 2005). Here we
present observations of ices toward three background stars over the
full 5-20 \micron\ range taken with the Infrared Spectrograph (IRS;
Houck et al. 2004) aboard {\it Spitzer}.  We assess the complete ice
inventory in quiescent clouds and compare it to observations toward
protostars.

\section{Observations and Reduction}

The background stars discussed here include a source behind the
Serpens dark cloud, CK 2, and two sources behind the Taurus dark
cloud, Elias 13 and Elias 16.  The observations are part of the
``c2d'' legacy program (Evans et al.\  2003). All sources were observed
with the short wavelength, low resolution module (SL; $\lambda =$ 5-14
\micron; $R \equiv \lambda / \Delta \lambda =$ 64-128), with on-source
integration times of 28 sec per spectral order.  CK 2 and Elias 13
were also observed with the short wavelength, high resolution module
(SH; $\lambda =$ 10-20 \micron; $R = 600$), with integration times of
240 sec and 60 sec per spectral order, respectively.  Spectra of
Elias 13, Elias 16 and CK 2 were part of AOR\# 0005636864, 0005637632,
and 0011828224. The SH spectrum of Elias 16 has been published by Bergin
et al.\ (2005) and was part of AOR\# 0003868160.
The data were
reduced using the {\it Spitzer} Science Center (SSC) pipeline version
S11.0.2 (S12 for SH in Elias 16) to produce the
2-dimensional Basic Calibrated Data.  Subsequently,
customized source extractions were performed, including the
subtraction of extended background emission. The spectra were then
defringed with sine-wave fitting routines (Lahuis \& Boogert 2003).

\begin{figure}
\epsfig{file=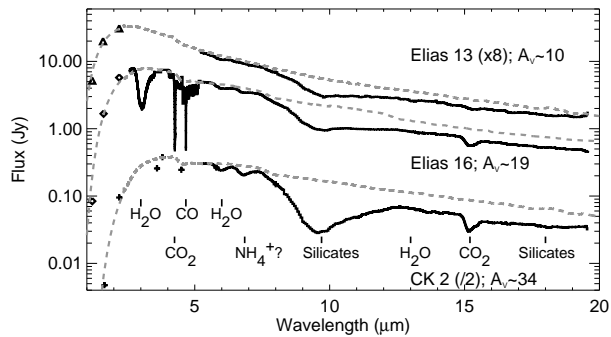,width =1.9in, angle=90} \caption{The spectra of
Elias 13, Elias 16 and CK 2 with the extincted photospheres (dashed
lines) used for normalization.  Near-infrared complementary data are
shown in triangles, diamonds and plus signs, respectively. The
spectra are scaled along the flux axis with the numbers given in
brackets. \label{fig:flux} }
\end{figure}

Figure {\ref{fig:flux}} shows the {\it Spitzer} spectra of the
observed background stars, complemented by near-infrared (NIR) broad band
photometry (2MASS\footnote{This publication makes use of data from the
Two Micron All Sky Survey, which is a joint project of the University
of Massachusetts and IPAC/Caltech, funded by NASA and NSF.}). In
addition, for CK 2, \spitzer\ IRAC (S. T. Megeath, priv. comm.) and
ground-based L$'$ band photometry (Churchwell \&\ Koornneef 1986) are
included and for Elias 16 the 2-5 \micron\ \iso\ spectrum is shown
(Whittet et al.\ 1998).

In order to put the data on an optical depth scale and analyze the
ice and dust features, each spectrum is normalized to the spectrum
of an extincted late-type giant taken from the \iso\ database (Sloan
et al.\ 2003).  A blackbody is used at wavelengths below 2.5
\micron\ to fit the NIR photometry. The extinction law used for
Serpens is of the form $A_{\lambda} \sim \lambda^{-1.9}$ (Kaas et
al.\ 2004), whereas that for Taurus has a shallower power law,
$A_{\lambda} \sim \lambda^{-1.7}$ (Whittet et al.\ 1988). Indebetouw
et al. (2005) find a shallower slope on the extinction curve beyond
6 \micron.  This flattening is partly due to the silicate and ice
features which we account for separately and the powerlaw is a good
approximation to the extinction at longer wavelengths. $A_{\rm V}$
is taken to be 9.1 $\times$ $A_{\rm K}$ (Rieke \&\ Lebofsky 1985).
Spectral types K4 III, G9 III, and K3 III were adopted for CK 2,
Elias 13 and Elias 16, respectively. These values are within the
range of types given in Chiar et al.\ (1994) for CK 2 and close to
the K2 III type assigned by Smith et al.\ (1993) for Elias 13 and
16.  While the selected spectral types give the best fit (in
removing the photospheric CO and SiO bands at 5 and 8 \micron), they
are uncertain by a few subclasses resulting in an abundance error of
20\% on the strong features, especially since the available \iso\
database has limited coverage.

\begin{figure}
\epsfig{file=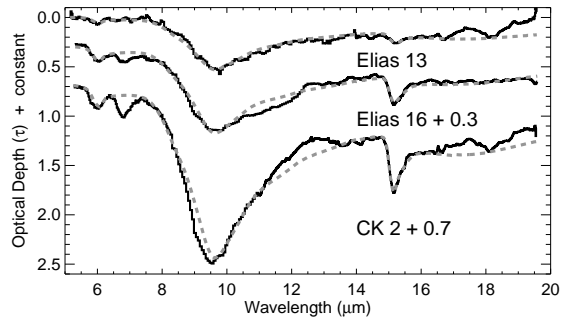, width=1.80in, angle=90} \caption{Background
star spectra on optical depth scale (black).  A fit (dashed grey) of
small spherical silicate grains is shown to which are added
laboratory spectra of \water\ and \coo\ (at 15 \micron) at
intensities corresponding to the column densities mentioned in Table
1. \label{fig:tauall}}
\end{figure}

\section{Results}

\subsection{\coo\ and \water\ Ices}

\begin{figure}
\epsfig{file=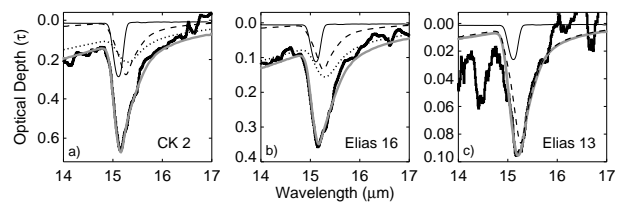, width=1.10in, angle=90} \caption{ \coo\ feature
for a) CK 2, b) Elias 16 and c) Elias 13 compared to laboratory
spectra composed of \water:\coo=10:1 (dot) and 1:1 (dash) at 10 K
and CO:\nn:\coo=100:50:20 (thin solid) at 30 K. The grey line shows
the best composite fit. Note that the vertical offset is due to the
\water\ libration mode. \label{fig:co2} }
\end{figure}

The \coo\ 15 $\mu$m feature is very strong toward CK 2 and Elias 16,
and weakly detected toward Elias 13.  The abundance of \coo\
relative to \water\ is 33\% toward CK 2, higher than the $\sim$20\%
abundance seen toward the Taurus sources.  The derived \coo\ column
densities toward Elias 13 and Elias 16 agree within errors with
those obtained from the 4.25 \micron\ feature (Nummelin et al.\
2001; Table 1). The bottom of the 15 \micron\ \coo\ feature appears
single peaked toward all sources (Fig. \ref{fig:co2}) and does not
show double dips due to crystallization as some protostars do
(Gerakines et al.\ 1999, Boogert et al.\ 2004).  The profile of this
band toward Elias 16 is fitted in Bergin et al.\ (2005) with the sum
of the polar (\water:\coo=7:1) and the apolar (CO:\coo=4:1)
mixtures. The polar, \water-rich mixture accounts for 85\% of the
CO$_2$ column density. However, the H$_2$O column density assumed in
this fit overestimates the observed value by 30\%. We require that
both the observed H$_2$O and CO$_2$ column densities, as well as the
CO$_2$ band profile are matched.  We use a combination of two polar
mixtures H$_2$O:CO$_2$=1:1 and H$_2$O:CO$_2$=10:1 (with the ratios
of the two mixtures: 2:1 for CK 2, 1.3:1 for Elias 16 and 1:0 for
Elias 13) and the apolar mixture CO:N$_2$:CO$_2$=100:50:20, all at
low temperature (Ehrenfreund et al. 1997). Satisfactory fits are
obtained for polar fractions of 78\% in CK 2, 84\% in Elias 16, and
87\% in Elias 13, comparable to the fractions found by Bergin et al
(2005) despite the different mixtures used.

\begin{figure}
\epsfig{file=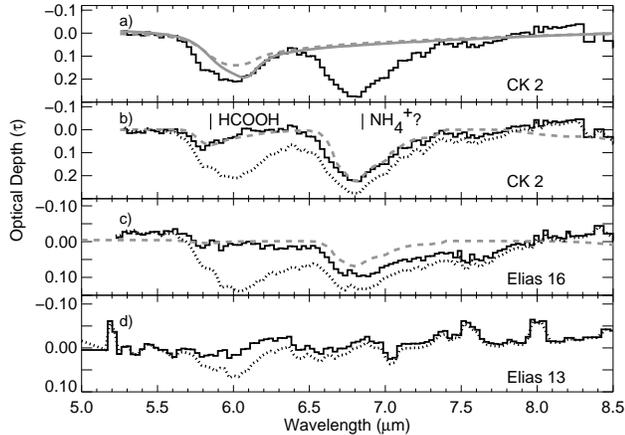, width=2.4in, angle=90}
\caption{a)The spectrum
of CK 2 after removing silicate absorption is shown in black. The
dash grey line is the pure \water-ice contribution to the 6.0
\micron\ feature while the solid grey is the \water:\coo\ mixture.
b-d) The dotted lines show the spectra after removing the silicate
contribution. The solid lines show the spectra after removing
silicate and \water\ contributions.  b) For CK 2, the dash grey line
shows a fit to HCOOH (5.85 and 8.2 \micron), \ammonium\ (6.85
\micron) and \ammonia\ (6 \micron).  c) For Elias 16, the dash grey
line shows a fit to \ammonium\ and \ammonia.  d) For Elias 13, no
remaining features are found.
\label{fig:otherices} }
\end{figure}

Using the \water\ ice column densities from the 3 \micron\ band
(Table 1) and laboratory spectra of pure \water\ ice (Hudgins et
al.\ 1993), the \water\ bending mode contributes to 77\% and 69\% of
the observed 6.0 \micron\ absorption feature for Elias 16 and CK 2
(Fig.~\ref{fig:tauall}). However, the peak position, width, and
strength of this band change significantly when \water\ is diluted.
For example, compared to pure \water, the mixture \water:\coo=1:1
shifts the peak to longer wavelengths by $\sim$0.1 \micron, and
increases the peak by a factor of 2.4, but the extensive long
wavelength wing remains unchanged. Using mixtures that fit the 15
\micron\ \coo\ band (Fig. {\ref{fig:co2}}), $\sim$85-100\% of the
6.0 \micron\ feature can be explained by \water\ (Fig.
{\ref{fig:otherices}}). The strong libration mode of \water\
explains much of the excess absorption in the 12-13 \micron\ region,
but due to severe blending with the silicate absorption feature,
residuals in that spectral region are hard to interpret (\S 3.3). A
model of astronomical silicates (Weingartner \&\ Draine 2001) is
used to fit the 10 and 20 \micron\ features for all sources (Fig.
{\ref{fig:tauall}}), but this silicate model may not be unique.

\subsection{The 6.85 \micron\ Band and Other Ices}

\begin{figure}
\epsfig{file=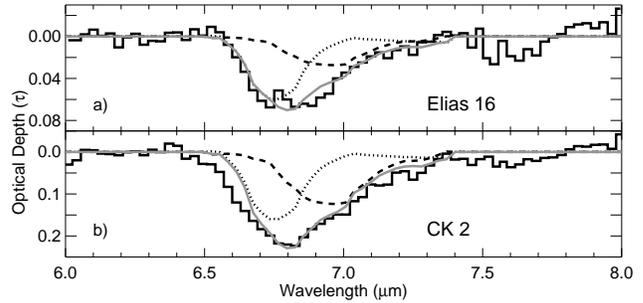, width=1.70in,angle=90} \caption{The
decomposition of the 6.85 \micron\ feature into short (dot) and long
wavelength (dash) components (Keane et al.\ 2001) for a) Elias 16
and b) CK 2. The solid grey line indicates the sum of the short and
long wavelength components. The solid black line represents the
spectrum after subtraction of silicate and \water\ contributions.
\label{fig:nh4p} }
\end{figure}

A strong feature at 6.85 \micron\ is detected toward CK 2 and Elias
16 but not toward Elias 13. This is the first detection of this band
toward background stars.  It is commonly observed toward protostars
(Keane et al.\ 2001) and often attributed to the \ammonium\ ion (see
Schutte \&\ Khanna 2003; \S 4). Regardless of the identification,
the profile may be a powerful tracer of the thermal history of the
ices.  Figure {\ref{fig:nh4p}} shows the decomposition of the 6.85
\micron\ feature into the short and long wavelength components used
by Keane et al.\ (2001) to characterize this band. The
phenomenological separation is meant to represent \ammonium\ bands
at different temperatures. The ratio of the peak optical depths of
the two components (short/long) is 2 for Elias 16 and 1.2 for CK 2.
If instead the equivalent widths are compared, the ratios are 1.6
and 0.9, respectively.  In either case, the profiles resemble those
of the protostars with the coldest sight-lines (e.g., NGC 7538
IRS9). Its equivalent width is used to calculate the column density
for \ammonium\ (Table 1) using the band strength from Schutte \&\
Khanna (2003).

\subsection{Other Ices and Hydrogen Column Densities}

\water\ ice accounts for up to $85\%$ of the 6.0 \micron\ absorption
band for CK 2 (Fig. {\ref{fig:otherices}}; \S 3.1).  Some excess
absorption remains, in particular on the short wavelength side. This
is observed towards protostars as well, and attributed in part to
absorption by HCOOH (Schutte et al.\ 1999, Keane et al.\ 2001).  The
5.85 \micron\ band is the strongest HCOOH band in our observed
wavelength range, that at 8.2 \micron\ is slightly weaker.  For CK
2, the laboratory spectrum of pure solid HCOOH fits the 5.85
\micron\ feature with a discrepancy at 8.2 \micron, which may
indicate over-correction of the photospheric SiO band (\S 2). Table
1 shows upper limits and tentative detections of weak features in
the 7--13 \micron\ spectral region.

\begin{deluxetable}{llccccc}
\tablewidth{0pt}
\tablecaption{Column Densities and Abundances }
\tablehead{ \colhead{Species}      & \colhead{Unit}      &
\colhead{CK 2}          & \colhead{EL 13}      & \colhead{EL 16}
& \colhead{HH46$^1$} & \colhead{B5$^1$} } \startdata
CO           & \%\water       & 36$^2$,      & 9$^4$  & 26$^4$              & 20    & 43\\
         &            & 57$^3$       &        &         &       &   \\
\coo         & \%\water       & 33           &  $\leq$15,   & 18$^5$,22$^6$, & 32    & 37\\
             &                &              & 22$^5$ & 19$^7$,24        &       &  \\
HCOOH        & \%\water       & 1.9          & ...          & ...                      &$<$8.7 & 9.3\\
\methanol    & \%\water       & $<$2.1       & $<$0.8       & $<$2.3                    & 7.0   & $<2.3$\\
\ammonia     & \%\water       & $\leq$8      & $<$6         & $\leq$8                   & 17    & $<$9.1\\
\ammonium    & \%\water       & 10.8         & ...          & 5.5                       & 9.6   & $<$12.7\\
\methane     & \%\water       & $<$3         & ...          & $<$3                      & 4     & ... \\
\ocn         & \%\water       & ...          & ...          & $<$2.3$^8$        & $\leq$0.7 & $<$0.5 \\
\\
\water       & 10$^{18}$ \cmc & 3.5$^9$ & 1.0$^{10}$        & 2.5$^{10}$       & 8.4   & 2.6\\
H            & 10$^{22}$ \cmc & 6.4          & 1.9          & 3.6                       & 5.0   & 3.2 
\enddata
\tablerefs{1.\ HH46 IRS and B5 IRS1 from Boogert et al.\ 2004, and
in preparation; 2.\ Chiar et al.\ 1994; 3.\ Pontoppidan, et al.\
2003; 4.\ Chiar et al.\ 1995; 5.\ Whittet et al.\ 1998, 6.\ Nummelin
et al.\ 2001; 7.\ Bergin et al.\ 2005, 8.\ Whittet et al.\ 2001; 9.\
Eiroa \&\ Hodapp 1989; 10.\ Whittet et al.\ 1988, Smith et al.\ 1993
}
\end{deluxetable}

The uncertainty in the shape of the silicate feature is large in
some places (e.g., 20$\%$ at 10--13 \micron) and hard to quantify in
others. For ice abundance determinations, the hydrogen column
density is usually calculated from the relation $N_{\rm H} \approx
1.87 \times 10^{21} \times A_{\rm V}$ cm$^{-2}$ for $R_{\rm V} =
3.1$ (Draine 2003). As outlined in \S 2, by fitting \iso\ template
spectra, extinctions \av\ of 34$^m$, 19$^m$, and 10$^m$ are derived
for CK 2, Elias 16, and Elias 13 respectively.  The observed peak
optical depth of the silicate band and the relation $A_{\rm
V}$/\tsil$= 18.5 \pm 2$ (Draine 2003) give similar values for \av.
The $N_{\rm H}$ values listed in Table 1 are calculated using the
\av\ values derived from the photospheric fits.

\section{Conclusions}

Our \spitzer\ spectra of background stars show clear detections of
absorption features at 6.0 and 6.85 \micron\ that had previously
only been seen toward YSOs, imposing new constraints on the origin
of the 6.85 \micron\ band.  The strength of the 6.85 \micron\ band,
scaled to \water, is similar to that seen toward YSOs, as is the
factor of 2 variation between sight-lines (Table 1; Schutte \&
Khanna 2003). In one scenario, the 6.85 \micron\ band is explained
by \ammonium\ produced by acid-base reactions in ice mixtures
containing \ammonia\ and HNCO. In laboratory experiments such
reactions occur at temperatures as low as 10 K with conversion
factors between 15\% and 100\% depending on ice mixtures and
temperature (van Broekhuizen et al. 2004). The strength of the 6.85
\micron\ band is a factor of 2 larger toward CK~2 compared to
Taurus. This sight line probes a very cold region, as evidenced by
the very high abundance of the volatile apolar CO ice ($T_{\rm
subl}\sim$20 K) as well as the smooth profile of the 15 \micron\
\coo\ band. In contrast, the line of sight of the YSO HH46 IRS shows
evidence for the ices to have undergone thermal processing (Boogert
et al.\ 2004), but the 6.85 \micron\ band is not unusually deep
(Table 1). Thus, if the 6.85 \micron\ band is due to \ammonium,
variables other than temperature, such as the initial HNCO and
\ammonia\ abundances, must play roles in determining its strength.
We stress that the identification of the 6.85 \micron\ band with
\ammonium\ is tentative and more evidence, including correlation
with the bands of counter ions such as \ocn\ (van Broekhuizen et
al.\ 2005) or HCOO$^-$ is required.

Most ($\sim$75\%) of the 6.0 \micron\ band toward background stars
is explained by pure \water\ ice, comparable to the percentage
toward most massive YSOs (Keane et al.\ 2001).  While \water\ mixed
with \coo\ can explain the remaining absorption in the Taurus
sources, the residual toward CK 2 has a peak absorption wavelength
consistent with HCOOH (Fig. {\ref{fig:otherices}}). Its abundance
would be a few \% of \water, comparable to that seen in several
high-mass YSOs (Keane et al.\ 2001) but not as high as seen in some
low-mass YSOs (Boogert et al., in preparation).

Dust grains have accumulated rather complex icy mantles in opaque
regions of molecular clouds before star formation begins, a point
which must be included in models of star formation.  Also, the
effect of freeze-out on the thermal balance has been studied by
Goldsmith (2001).  For the two stars with extinctions above 15 mag,
the abundances relative to \water\ ice are within the range seen
toward embedded objects. From the abundances in Table 1, the
percentages of nitrogen, oxygen and carbon locked in ices for Elias
16 and CK 2 are 35-37\% N, 28-30\% O and $\sim$12\% C. Further work
on larger samples of background stars will elucidate the dependence
of the ice composition on cloud conditions and history. Such surveys
are now possible with \spitzer/IRS for sources as weak as 10 mJy
(9.5 mag) at 8 \micron\ and \av\ of up to 50 mag.

\acknowledgements{The authors thank S. T. Megeath and H.\ Fraser for input.
Support for this work, part of the \spitzer\
Legacy Science Program, was provided by NASA through contracts
1224608, 1230779, 1256316, issued by the Jet Propulsion Laboratory,
California Institute of Technology, under NASA contract 1407.
Astrochemistry in Leiden is supported by a NWO Spinoza and NOVA grant,
and by the EU RTN-PLANETS (HPRN-CT-2002-00308).}

\end{document}